\documentclass{siam-wns-article}

\usepackage{amsmath}
\usepackage{tabularx}
\usepackage{graphicx}
\usepackage{subfigure}
\usepackage{amssymb}
\usepackage{amsthm}
\usepackage{algorithm}
\usepackage{algpseudocode}

\title{Optimal Control of Interdependent Epidemics in Complex Networks}

\author{Juntao Chen, Rui Zhang, Quanyan Zhu}

\usehyperref 

\begin{document}
\maketitle

\section*{Summary}
Optimal control of interdependent epidemics spreading over complex networks is a critical issue. We first establish a framework to capture the coupling between two epidemics, and then analyze the system's equilibrium states by categorizing them into three classes, and deriving their stability conditions. The designed control strategy globally optimizes the trade-off between the control cost and the severity of epidemics in the network. A gradient descent algorithm based on a fixed point iterative scheme is proposed to find the optimal control strategy. The optimal control will lead to switching between equilibria of the interdependent epidemics network. Case studies are used to corroborate the theoretical results finally.    
\section{Introduction}
Control of epidemics in complex networks is a prevailing problem ranging from social science to engineering \cite{hansen2011optimal,newman2002spread}. A network containing two interdependent epidemics with a control $\textbf{u}:=(u_1,u_2)\in\mathbb{R}_+^2$ can be described by a model similar to the one in \cite{pastor2001epidemic}:
\begin{equation}\label{strains}
\begin{split}
\frac{dI_{1,k}(t)}{dt}=-\gamma_1 I_{1,k}(t)
&+\zeta_1k[1-I_{1,k}(t)\\
&-I_{2,k}(t)]\Theta_1(t)
-u_1I_{1,k}(t),\\
\frac{dI_{2,k}(t)}{dt}=-\gamma_2 I_{2,k}(t)
&+\zeta_2k[1-I_{1,k}(t)\\
&-I_{2,k}(t)]\Theta_2(t)
-u_2I_{2,k}(t),
\end{split}
\end{equation}
where $I_{1,k}(t)$ and $I_{2,k}(t)$ represent the densities of nodes at time $t$ with degree $k$ infected by virus strain 1 and strain 2, respectively; $(\gamma_1,\gamma_2)$ and $(\zeta_1,\zeta_2)$ are recovery and spreading rates of two strains;
$
\Theta_1(t)=\frac{\sum_{k'}k'P(k')I_{1,k'}(t)}{
\langle k \rangle},
\Theta_2(t)=\frac{\sum_{k'}k'P(k')I_{2,k'}(t)}{
\langle k \rangle},
$ where $P(k)$ is the probability distribution of a node with degree $k$, and $\langle k \rangle=\sum_k kP(k)$.

The network cost over a time period $[0,T]$ is captured by two terms: the control cost $c_1(\textbf{u})$ and the severity of epidemics $c_2(\bar{I}_1(t)+\bar{I}_2(t))$, where $c_1$ and $c_2$ are both monotonically increasing functions. In addition,
$
\bar{I}_1(t):=\sum_k P(k)I_{1,k}(t)$ and 
$
\bar{I}_2(t):=\sum_k P(k)I_{2,k}(t),
$
and they can be interpreted as the severity of epidemics in the network.
The optimal control problem at network equilibrium can be formulated as
\begin{equation*}
\begin{split}
(\mathrm{OP1}):\quad&\min_{\textbf{u}}\ c_1(\textbf{u})+c_2\big(\bar{I}_1^*(u_1)+\bar{I}_2^*(u_2)\big)\\
&\hspace{4ex}\mathrm{s.t.}\quad \mathrm{system\ dynamics}\ \eqref{strains},
\end{split}
\end{equation*}
where $\bar{I}_1^*(u_1)$ and $\bar{I}_2^*(u_2)$ denote the densities of the strains at the steady state under the control $\textbf{u}$. Note that (OP1) can also be interpreted as the average cost minimization problem. At the steady state, $dI_{1,k}/dt=0$ and $dI_{2,k}/dt=0$,  and we obtain
$
I_{1,k}=\frac{\psi_1k\Theta_1}{1+\psi_1k\Theta_1+\psi_2k\Theta_2}$ and
$
I_{2,k}=\frac{\psi_1k\Theta_1}{1+\psi_1k\Theta_1+\psi_2k\Theta_2},
$ where $\psi_i=\zeta_i/(\gamma_i+u_i),\ i=1,2$.
Then, the control problem (OP1) can be reformulated as
\begin{equation*}
\begin{split}
(\mathrm{OP2}):\quad &\min_{\textbf{u}}\quad c_1(\textbf{u})+c_2\big(\bar{I}_1^*(u_1)+\bar{I}_2^*(u_2)\big)\\
\mathrm{s.t.}\quad &{I}_{i,k}^*(u_i)=\frac{\psi_ik\Theta_i^*}{1+\psi_ik\Theta_i^*+\psi_{-i}k\Theta_{-i}^*},\ i=1,2,
\end{split}
\end{equation*}
where $-i=\{1,2\}\setminus \{i\}$, $\Theta_i^*=\frac{\sum_{k'}k'P(k')I_{i,k'}^*(u_i)}{
\langle k \rangle}$, and
$\bar{I}_i^*(u_i)=\sum_k P(k)I_{i,k}^*(u_i)$ is the total number of nodes infected by strain $i$.

Our objective is to design a control strategy via solving (OP2) which jointly optimizes the control cost and the epidemics spreading level in the network.

 \begin{figure*}[t]
  \centering  
  \subfigure[optimal control ($E_1$)]{%
    \includegraphics[width=0.25\textwidth]{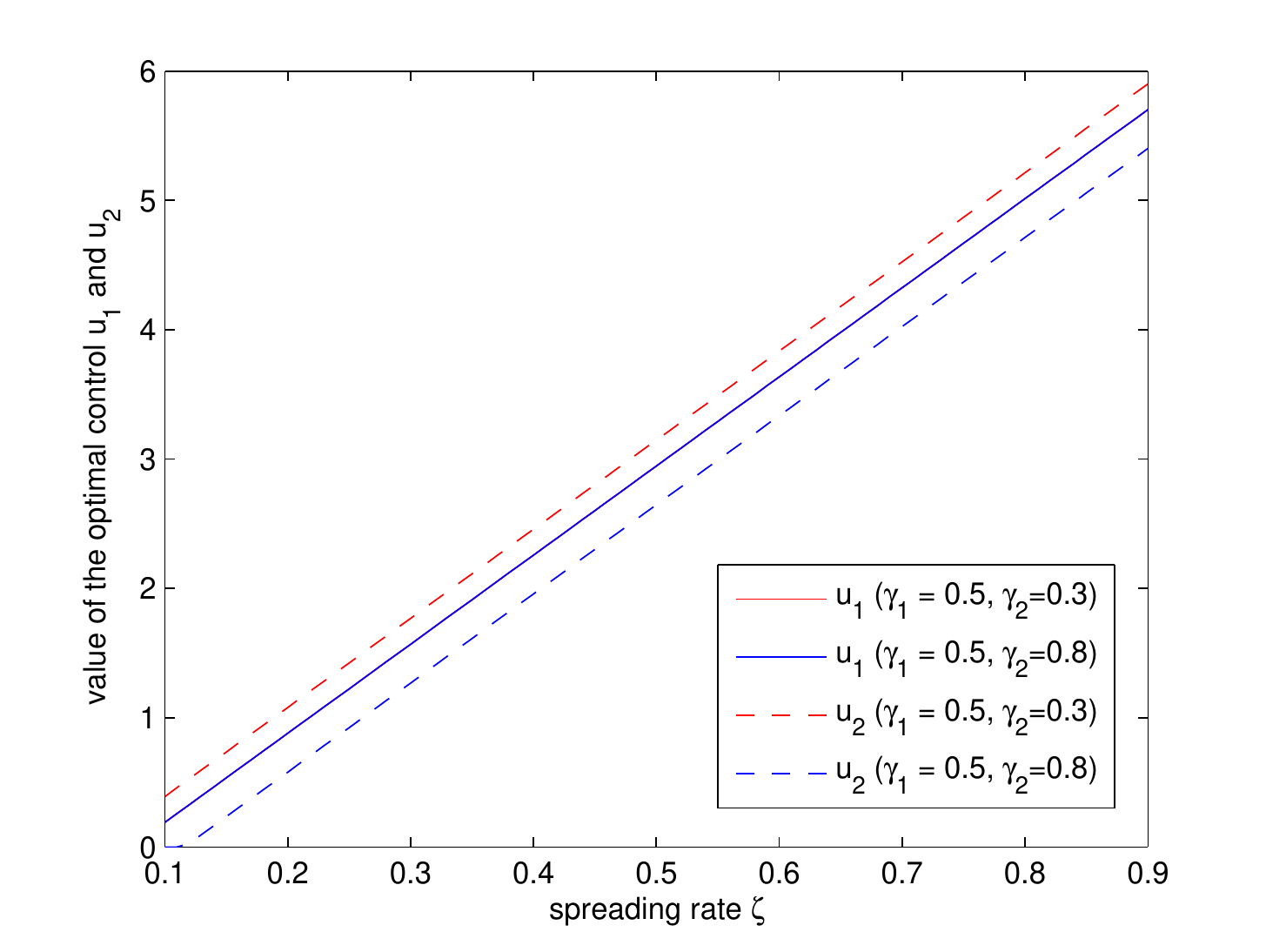}%
    \label{fressresults}%
  }%
  \hfill
  \subfigure[objective ($E_1$)]{%
    \includegraphics[width=0.25\textwidth]{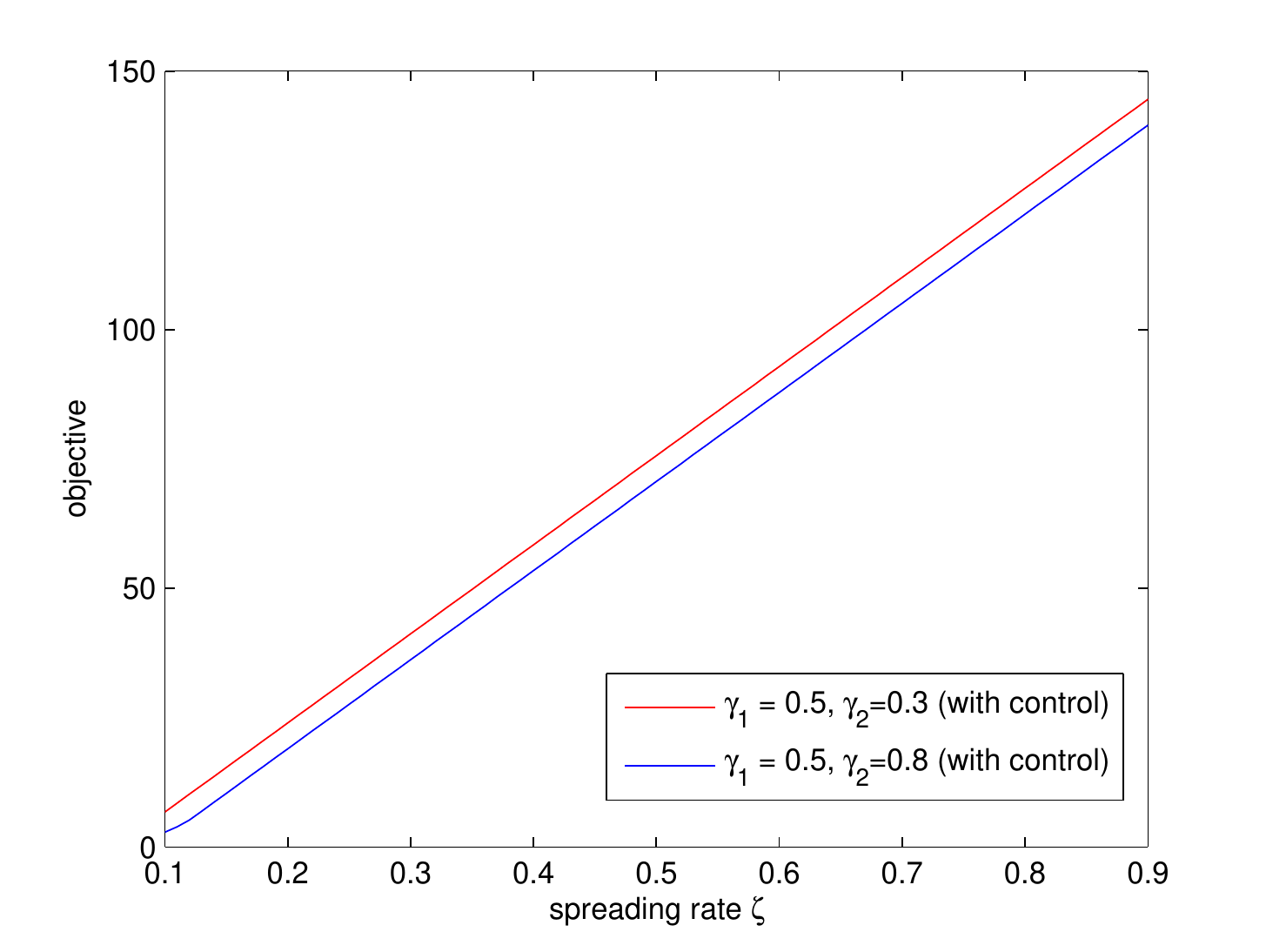}%
    \label{fressresults2}%
  }%
  \hfill
  \subfigure[optimal control strain 1 ($E_2$)]{%
    \includegraphics[width=0.25\textwidth]{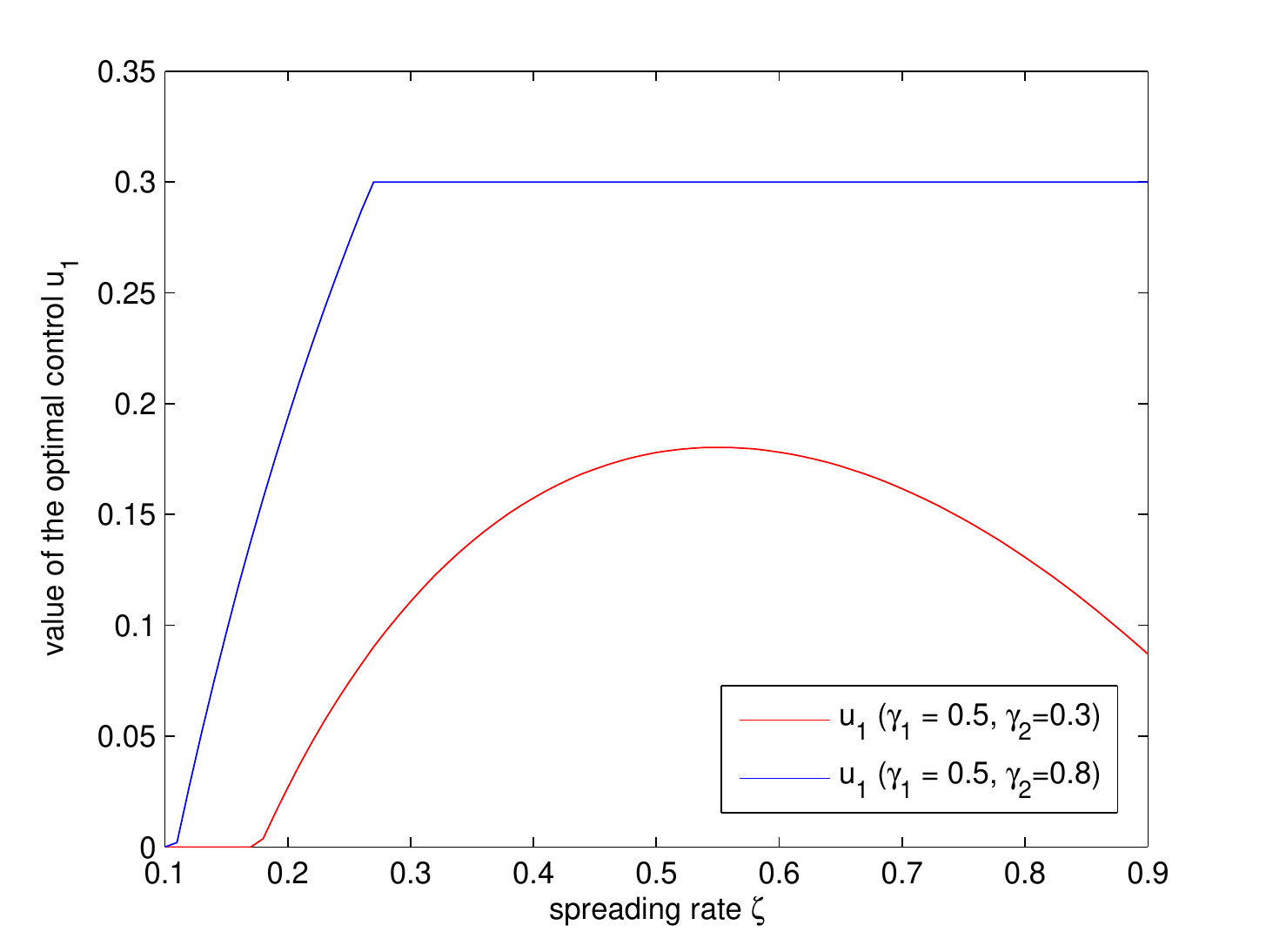}%
    \label{strain1_result}%
  }%
   \hfill
  \subfigure[optimal control strain 2 ($E_2$)]{%
    \includegraphics[width=0.25\textwidth]{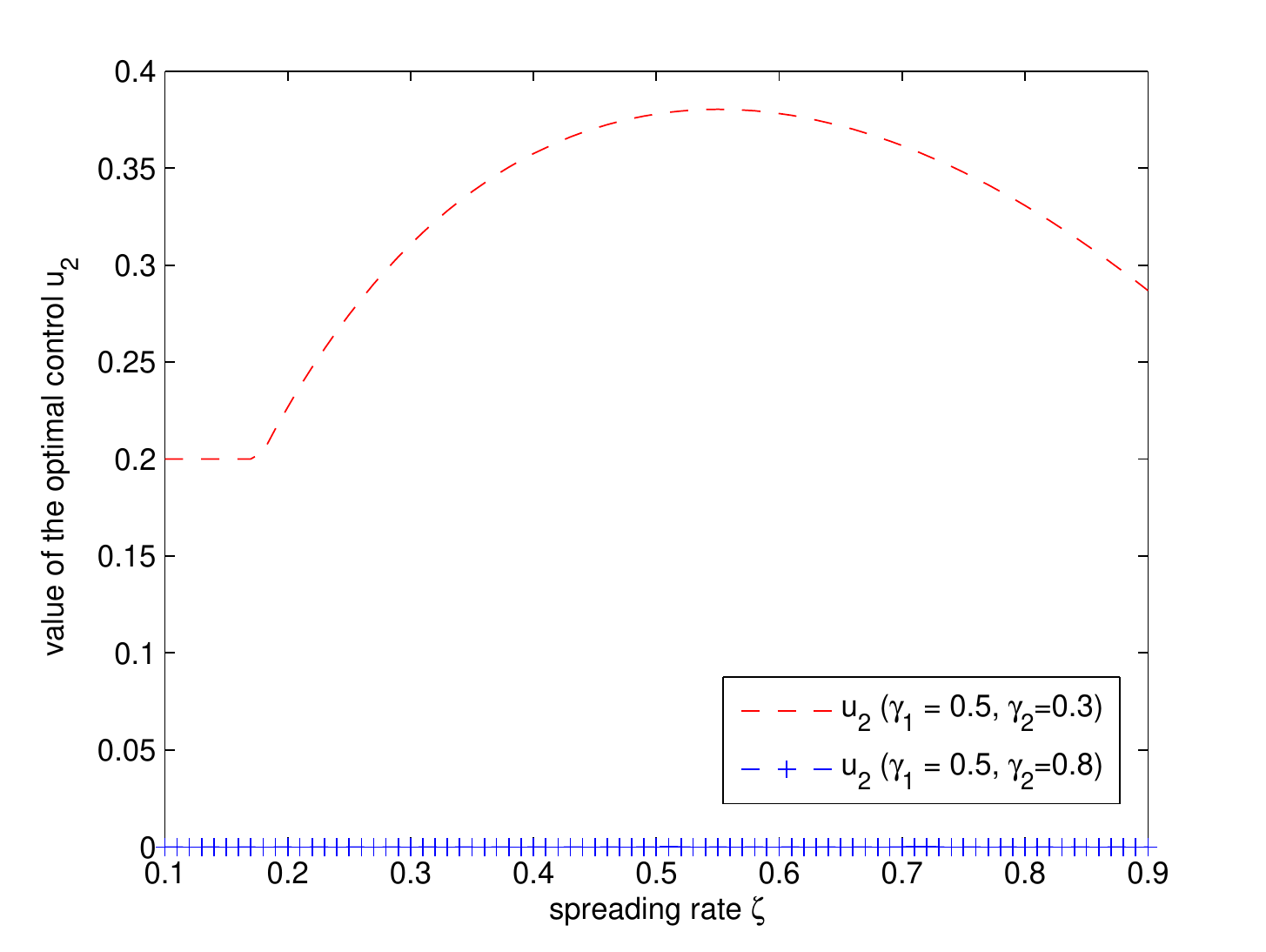}%
    \label{strain1_result2}%
  }%
     \hfill
  \subfigure[optimal control strain 1 ($E_3$)]{%
    \includegraphics[width=0.25\textwidth]{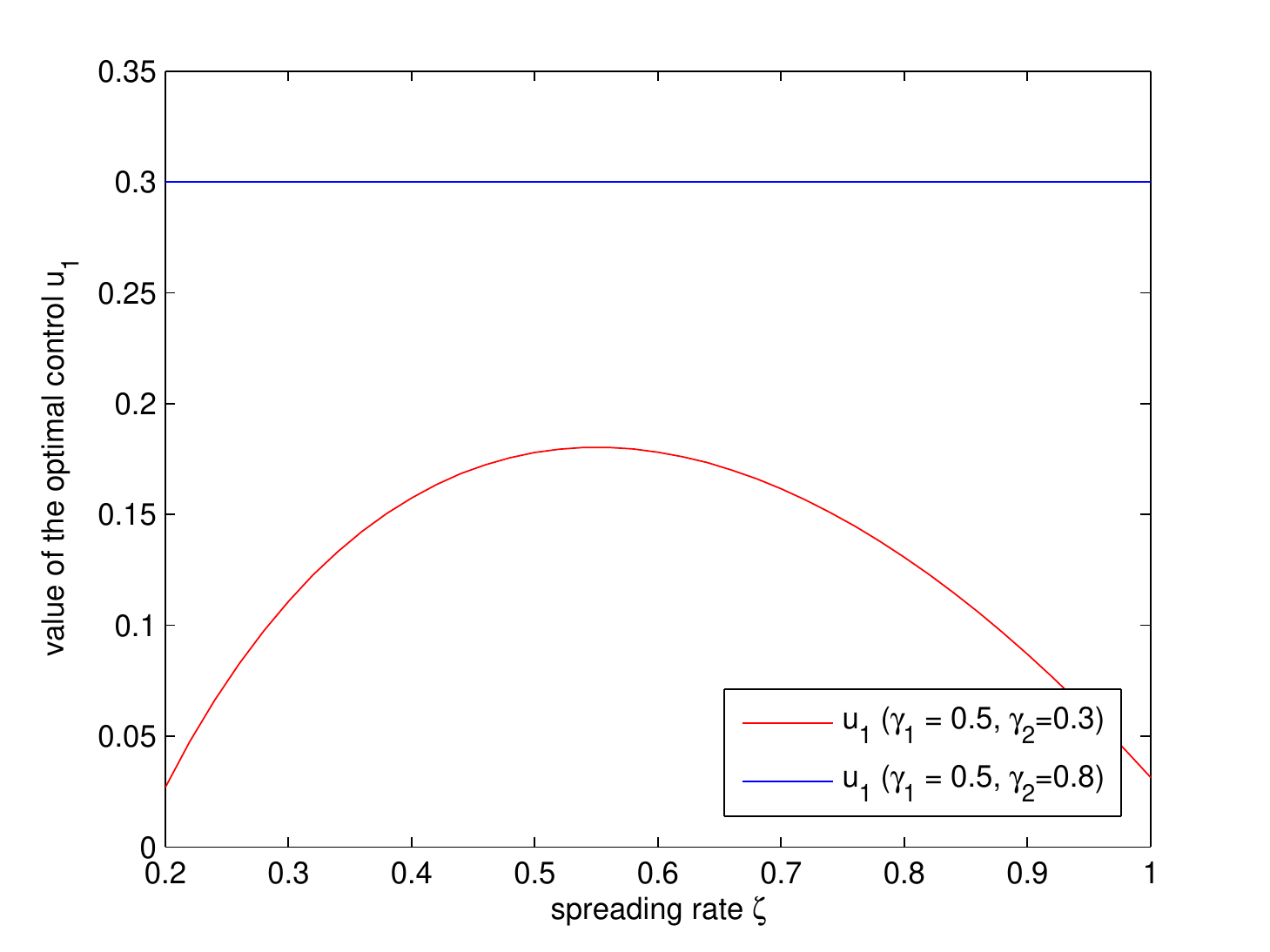}%
    \label{strain2_result}%
  }%
     \hfill
  \subfigure[optimal control strain 2 ($E_3$)]{%
    \includegraphics[width=0.25\textwidth]{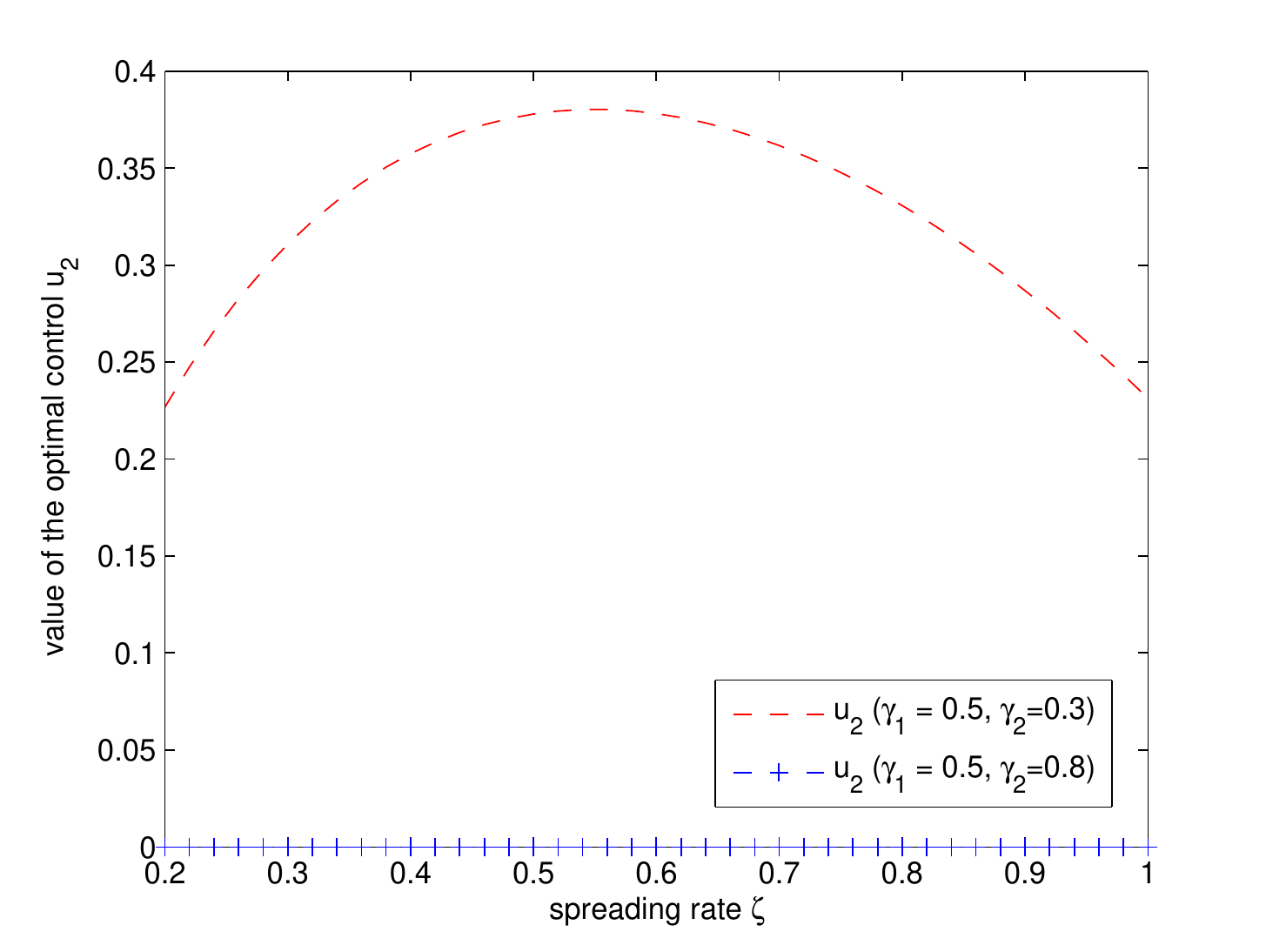}%
    \label{strain2_result2}%
  }%
     \hfill
  \subfigure[single switching case]{%
    \includegraphics[width=0.23\textwidth]{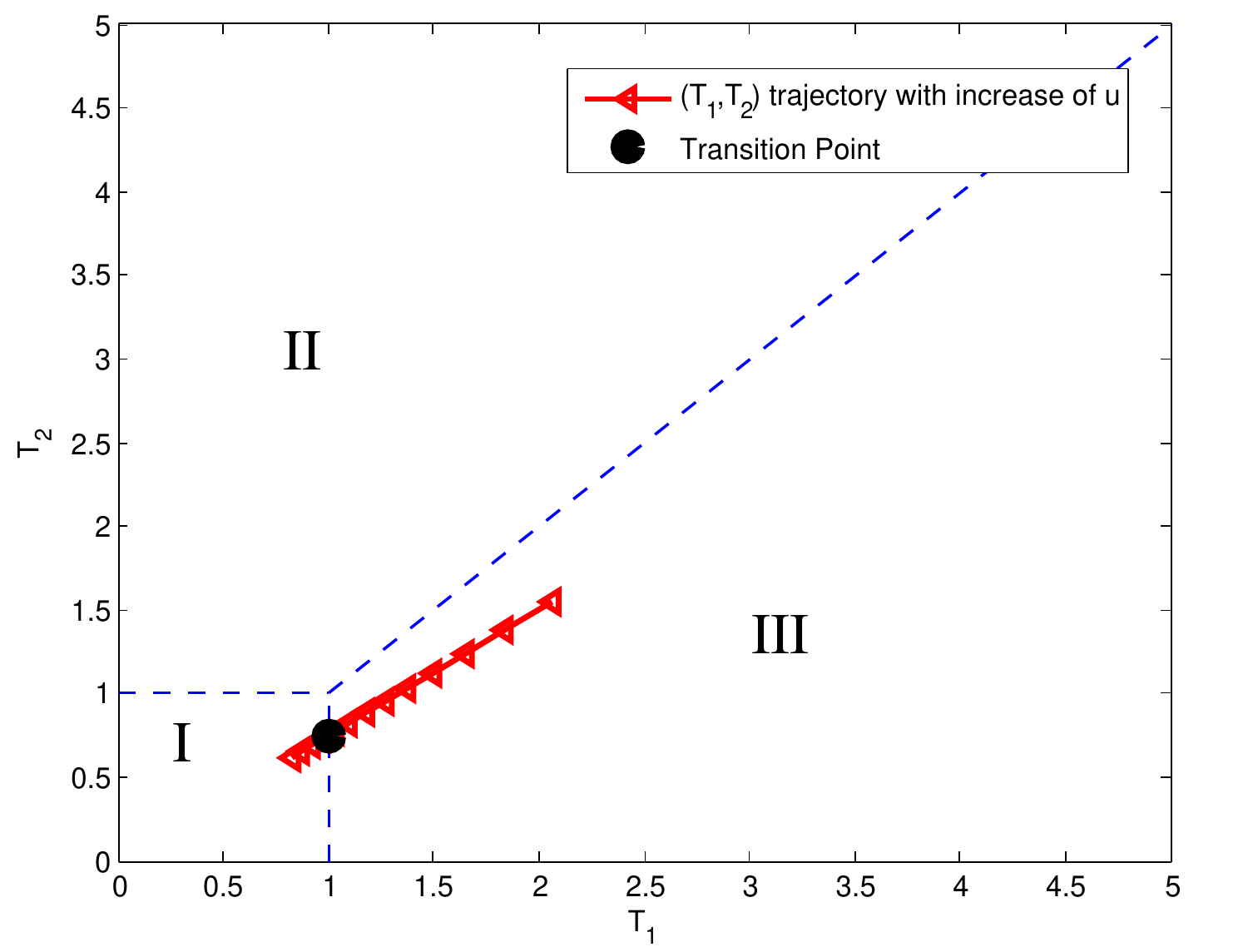}%
    \label{transition1}%
  }%
     \hfill
  \subfigure[double switching case]{%
    \includegraphics[width=0.23\textwidth]{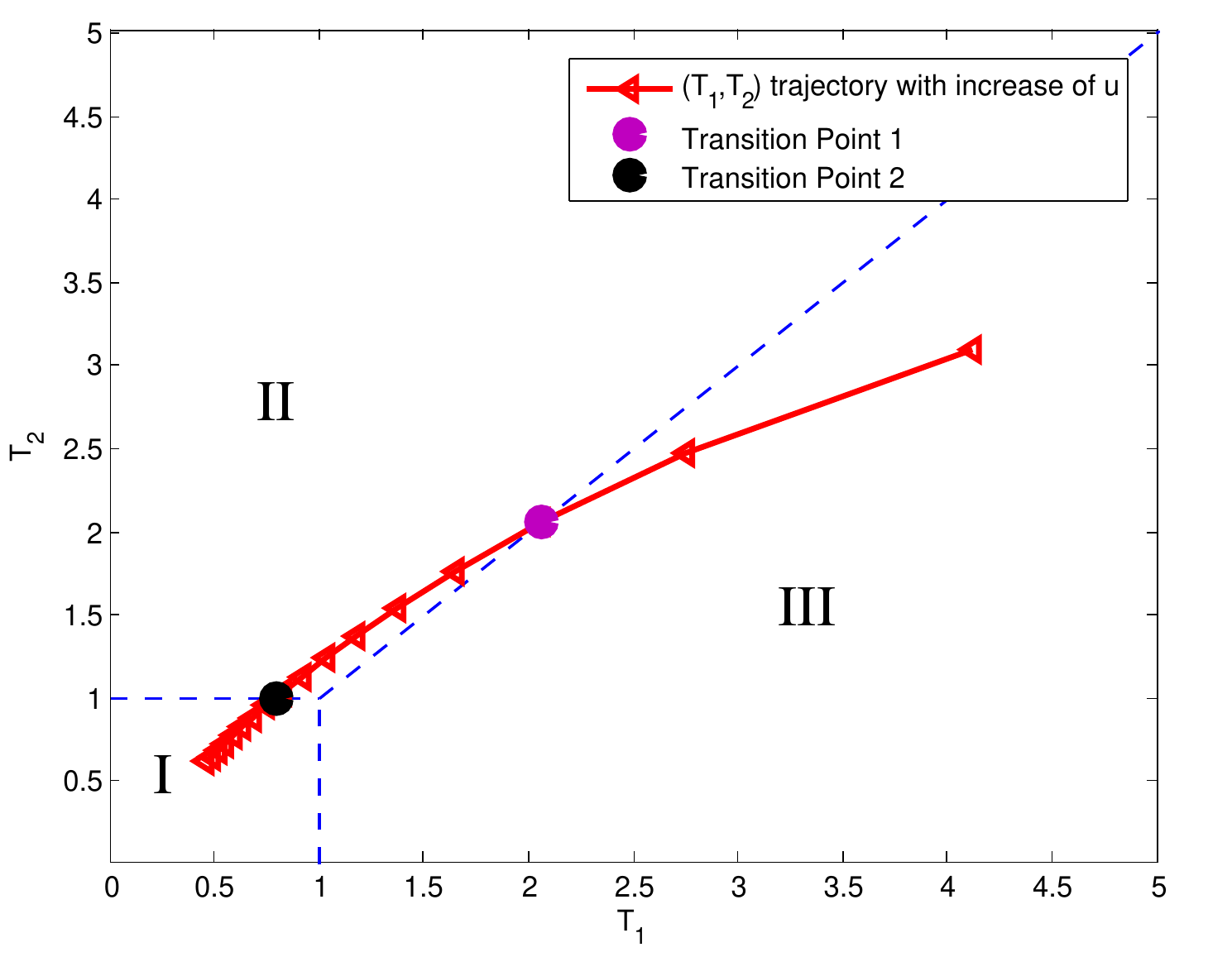}%
    \label{transition2}%
  }%
  \vspace{-2mm} \caption{Results of the optimal control for each equilibrium case, and the demonstration of switching of equilibrium.}
  \label{bigfig2}\vspace{-3mm}
\end{figure*}

\section{Main Results}
To solve (OP2), we first need to analyze the system's steady states. The equilibrium pair $(\Theta_1^*,\Theta_2^*)$ needs to satisfy the following self-consistency equations for $i=1,2$:
\begin{equation}\label{self1}
\Theta_i=\frac{\psi_i}{\langle k \rangle}\sum_{k'}\frac{k'^2P(k')\Theta_i}{1+\psi_ik'\Theta_i+\psi_{-i}k'\Theta_{-i}}.
\end{equation}
For equation \eqref{self1}, $(\Theta_1,\Theta_2)=(0,0)$ is an obvious solution that results in $\bar{I}_1^*=\bar{I}_2^*=0$ and leads to an epidemics-free equilibrium.
By a closer checking of \eqref{self1}, we conclude that there exist no positive solutions, i.e., $\Theta_1>0$ and $\Theta_2>0$. Hence, besides the epidemics-free one, the system has another two exclusive equilibria, which lead to either entire population infected by strain 1 or by strain 2.
The conditions that lead to different network equilibria are essential. Let $T_1:=\frac{\psi_1\langle k^2 \rangle}{\langle k \rangle}$, $T_2:=\frac{\psi_2\langle k^2 \rangle}{\langle k \rangle}$, and then, three possible equilibrium states can be summarized as follows: \textbf{(i)}: Epidemics-free equilibrium $\boldsymbol{E_1}$; \textbf{(ii)}: Exclusive equilibrium of strain 1, $\boldsymbol{E_2}$, if and only if $T_1>1$; \textbf{(iii)}: Exclusive equilibrium of strain 2, $\boldsymbol{E_3}$, if and only if $T_2>1$.

{\bf Stability Analysis}: Through an eigenvalue analysis of the nonlinear dynamic system \eqref{strains}, we obtain the following results. \textbf{(i)}: $E_1$ is asymptotically stable if and only if $T_1\leq 1$ and $T_2\leq 1$. \textbf{(ii)}: $E_2$ is asymptotically stable if and only if $T_1>1$ and $T_1>T_2$. \textbf{(iii)}: $E_3$ is asymptotically stable if and only if $T_2>1$ and $T_2>T_1$. 

{\bf Optimal Control}: For each case, we can further obtain its corresponding control bounds. Then, the optimization problem (OP2) can be simplified by dividing it into three stable equilibrium cases. For example, under $E_2$, i.e., when $\bar{I}_{2,k}^*=0$, (OP2) becomes $\min_{\textbf{u}}c_1(\textbf{u})+c_2\big(\bar{I}_1^*(u_1)\big)$ with constraints ${I}_{1,k}^*(u_1)=\frac{\psi_1k\Theta_1^*}{1+\psi_1k\Theta_1^*}$, $u_1<\frac{\zeta_1 \langle k^2 \rangle}{\langle k \rangle}-\gamma_1$ and $
u_2> \frac{\zeta_2(\gamma_1+u_1)}{\zeta_1}-\gamma_2$. By addressing the coupling terms $\bar{I}_{1,k}^*(u_1)$ and $\Theta_1^*$, we obtain a fixed point equation as
$
\Theta_1^*=\frac{1}{\langle k \rangle}\sum_{k'}\frac{k'^2P(k')\psi_1\Theta_1^*}{1+\psi_1k'\Theta_1^*}.
$
We can show that there exists a unique solution $\Theta_1^*$ to the fixed point equation, and also the mapping $u_1\rightarrow \bar{I}_1^*(u_1)$ is continuous. The solution $\Theta_1^*$ with respect to $\psi_1$ can be obtained via a fixed point iterative scheme of which the stability and convergence are guaranteed due to the contraction mapping. With obtained $\Theta_1^*$, the objective function is only related to $\textbf{u}$ and can be solved by the gradient descent method. Optimal control is achieved until both $\Theta_1$ and $\textbf{u}$ converge. Other equilibrium cases can be analyzed in a similar way.

Another finding is that when the equilibrium state of the network without control is not epidemics-free, then it can switch to different states with the increase of control effort. Depending on the parameters of the epidemics, the control can lead to either single or double switching between equilibrium points (see Figs. \ref{transition1} and \ref{transition2}).

{\bf Numerical Experiments}: Case studies based on a scale-free network are to validate the theoretical results. Specifically, the cost functions are chosen as $c_1=15u_1+10u_2$ and $c_2=50(\bar{I}_1^*(u)+\bar{I}_2^*(u))$. Strain 1 and strain 2 have the same spreading rate $\zeta_1=\zeta_2$. For comparison, we have two cases: (1) $\gamma_1=0.5,\gamma_2=0.3$ and  (2) $\gamma_1=0.5,\gamma_2=0.8$.
For the epidemics-free case, the results are shown in Figs. \ref{fressresults} and \ref{fressresults2}. In addition, the results corresponding to the exclusive equilibrium of strain 1 and strain 2 are shown in Figs. \ref{strain1_result} and \ref{strain1_result2} and Figs. \ref{strain2_result} and \ref{strain1_result2}, respectively. To demonstrate the switching of equilibria through control, we choose two cases: (1) $\zeta_1=0.2,\gamma_1=0.4,\zeta_2=0.15$ and $\gamma_2=0.4$; (2) $\zeta_1=0.1,\gamma_1=0.1,\zeta_2=0.15$, and $\gamma_2=0.2$. The obtained results are shown in Figs. \ref{transition1} and \ref{transition2}.

\bibliographystyle{abbrv}
\bibliography{wns-bib}

\end{document}